%% file: main.tex
\documentclass[sigconf,nonacm]{acmart}

\usepackage{tcolorbox}
\usepackage{fontawesome5}
\usepackage{todonotes}
\usepackage{pgfplots}
\usepackage{subcaption}
\usepackage{longtable}
\usetikzlibrary{
    positioning,
    arrows,
    calc,
    shapes.callouts,
    decorations.pathmorphing,
    decorations.text,
    fit,
    backgrounds,
}
\usepgfplotslibrary{statistics}
\pgfplotsset{compat=1.15}
\definecolor{c1}{HTML}{e41a1c}
\definecolor{c2}{HTML}{377eb8}
\definecolor{c3}{HTML}{4daf4a}
\definecolor{c4}{HTML}{984ea3}

\newcommand{\name}{\textsc{DL-MIA}}

\newcommand{\ms}{\textsc{MS MARCO}}
\newcommand{\trecdl}{\textsc{TREC-DL}}
\newcommand{\trecweb}{\textsc{TREC-Web}}
\newcommand{\codec}{\textsc{CODEC}}
\newcommand{\dlhard}{\textsc{DL-HARD}}
\newcommand{\clueweb}{\textsc{ClueWeb}}

\newcommand{\bm}{\textsc{BM25}}
\newcommand{\bert}{\textsc{BERT}}
\newcommand{\obm}{\textsc{BM25}}
\newcommand{\col}{\textsc{ColBERTv2}}

\newcommand{\otree}{\textsc{oTree}}

\copyrightyear{2024}
\acmYear{2024}
\setcopyright{rightsretained}
\acmConference[CIKM '24]{Proceedings of the 33rd ACM International Conference on Information and Knowledge Management}{October 21--25, 2024}{Boise, ID, USA}
\acmBooktitle{Proceedings of the 33rd ACM International Conference on Information and Knowledge Management (CIKM '24), October 21--25, 2024, Boise, ID, USA}
\acmDOI{10.1145/3627673.3679166}
\acmISBN{979-8-4007-0436-9/24/10}

\makeatletter
\gdef\@copyrightpermission{
 \begin{minipage}{0.3\columnwidth}
 \href{https://creativecommons.org/licenses/by/4.0/}{\includegraphics[width=0.90\textwidth]{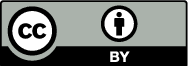}}
 \end{minipage}\hfill
 \begin{minipage}{0.7\columnwidth}
 \href{https://creativecommons.org/licenses/by/4.0/}{This work is licensed under a Creative Commons
Attribution International 4.0 License.}
 \end{minipage}
 \vspace{5pt}
}
\makeatother
\begin{document}
\title{Understanding the User: An Intent-Based Ranking Dataset}

\author{Abhijit Anand}
\affiliation{
  \institution{L3S Research Center}
  \city{Hannover}
  \country{Germany}
}
\email{aanand@L3S.de}

\author{Jurek Leonhardt}
\affiliation{
  \institution{Delft University of Technology}
  \city{Delft}
  \country{The Netherlands}
}
\email{L.J.Leonhardt@tudelft.nl}

\author{Venktesh V}
\affiliation{
  \institution{Delft University of Technology}
  \city{Delft}
  \country{The Netherlands}
}
\email{v.Viswanathan-1@tudelft.nl}
 
\author{Avishek Anand}
\affiliation{
  \institution{Delft University of Technology}
  \city{Delft}
  \country{The Netherlands}
}
\email{avishek.anand@tudelft.nl}

\begin{abstract}
As information retrieval systems continue to evolve, accurate evaluation and benchmarking of these systems become pivotal. Web search datasets, such as \ms{}, primarily provide short keyword queries without accompanying intent or descriptions, posing a challenge in comprehending the underlying information need. This paper proposes an approach to augmenting such datasets to annotate informative query descriptions, with a focus on two prominent benchmark datasets: \trecdl{}-21 and \trecdl{}-22.
Our methodology involves utilizing state-of-the-art LLMs to analyze and comprehend the implicit intent within individual queries from benchmark datasets. By extracting key semantic elements, we construct detailed and contextually rich descriptions for these queries. To validate the generated query descriptions, we employ crowdsourcing as a reliable means of obtaining diverse human perspectives on the accuracy and informativeness of the descriptions. This information can be used as an evaluation set for tasks such as ranking, query rewriting, or others. 
\end{abstract}


\maketitle
\input{intro-v2}
\input{rel_work}
\input{dataset_creation}
\input{experimental_setup}
\input{conclusion}


\bibliographystyle{ACM-Reference-Format}
\bibliography{bib}

\newpage
\clearpage
\appendix
\input{appendix}

\end{document}

%% file: intro-v2.tex
\section{Introduction}
\label{sec:intro}

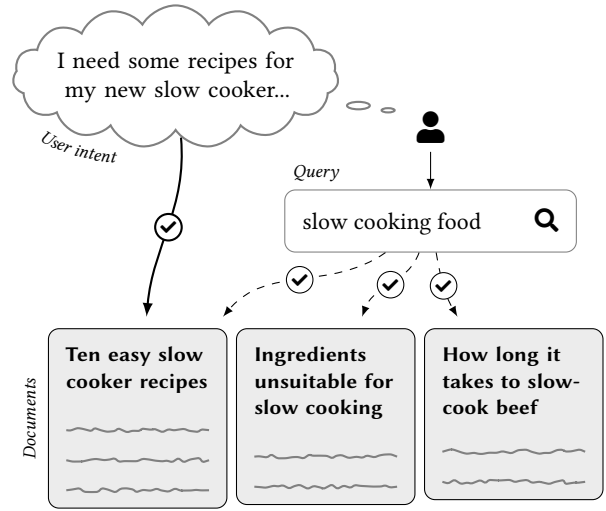
\begin{figure}
    \centering
    \input{fig/query-intent}
    \caption{An illustration of a user querying a search engine. The user has a specific intent in mind, but formulates the query in a more ambiguous way. As a result, there is a discrepancy between the documents relevant to the query and the documents relevant to the actual user intent.}
    \label{fig:intro.query_intent}
\end{figure}

In information retrieval (IR), a core challenge in building ranking models is to explicitly or implicitly \textit{aligning} the actual user intent with the machine intent, i.e., the intent as understood by the ranker.
This misalignment stems from the inherent complexity and variability in how users articulate their information needs versus how these needs are interpreted and processed by retrieval systems. 
This misalignment might be due to multiple reasons -- ambiguity, poorly formulated queries, complex queries, or a retrieval set that lacks relevant documents~\cite{mackie2021deep,collins2014trec}. 

Most current research on ranking models in IR is based on training parameterized models over large training datasets from \ms{}~\cite{nguyen2016ms}. 
However, to the best of our knowledge, there exist no recent datasets that attempt to measure the chasm between user intent and machine intent.
The current practice of measuring ranking performance is through sparsely~\cite{nguyen2016ms} or densely annotated ad-hoc ranking test sets~\cite{craswell2020overview,craswell2021overview,craswell2022overview,craswell2023overview,craswell2024overview} that provide queries and corresponding relevance annotations.
While these test sets allow for determining the overall effectiveness of a ranker, they fail to provide a way of measuring the extent to which the ranking models understand the true intent of the user.
For example, consider the query ``\textit{what are the three countries in 1984}''. 
While the intent—to identify the three countries mentioned in George Orwell's novel ``1984''—seems clear, it remains difficult to rank effectively because it requires specific contextual knowledge that may not be directly available in the retrieved documents. 
Another example is the query ``\textit{slow cooking food}'' (cf.\ Fig.~\ref{fig:intro.query_intent}).
Although this query appears to be straightforward, it can have multiple intents. 
This multiplicity of potential intents complicates the ranking process, as the system needs to correctly infer and prioritize the user’s actual intent to provide relevant results.
Knowing the user's intent allows the model to retrieve and rank documents most relevant to that intent, thereby addressing a critical challenge in handling ambiguous queries.

In this paper, we specifically focus on a subset of these challenges: \textit{queries that contain multiple intents}. 
We propose a new dataset named \name{} (\textbf{M}S MARCO \textbf{I}ntent \textbf{A}nnotations), which is a derivative of the \trecdl{} test sets.
The \name{} dataset contains 2655 tuples of \texttt{(query, intent, passage, label)} over a small yet challenging set of 24 queries from the \trecdl{} '21 and '22 datasets.
To construct \name{}, the key challenge was to accurately formulate user intents, as only queries are available in the \trecdl{} test sets. 
Toward this, we used a combination of LLM-generated query-specific intents and sub-intents that are post-processed through a carefully designed crowd-sourcing process to ensure human supervision and quality control.
\name{} mainly aims at measuring the gap between user intent and query by \textit{fine-grained intent annotation}, but can be used in multiple ranking scenarios, such as re-ranking, diversification, intent coverage, or query suggestion tasks. 

Our contributions are twofold -- first, we introduce a comprehensive dataset \name{} that meticulously documents the variations and complexities of user intent; second, we provide an analysis of this dataset's impact on ranking performance by applying it to several baseline models.
\name{} is publicly available at \url{https://zenodo.org/doi/10.5281/zenodo.11471482}.

%% file: fig/query-intent.tex
\tikzset{
    docheader/.style={
            align=left,
            text width=6em,
            inner sep=0,
            outer sep=0,
            font=\small\bfseries\sffamily,
        },
    relevant/.style={
            circle,
            draw=black,
            solid,
            fill=white,
            inner sep=0.1em,
            font=\scriptsize,
            execute at begin node=\faIcon{check},
        },
    qdarrow/.style={
            dashed,
        },
    faketext/.style={
            decorate,
            decoration={
                    random steps,
                    amplitude=0.1em,
                    segment length=0.25em,
                },
            gray,
            thick,
            rounded corners=0.075em,
        },
}

\begin{tikzpicture}
    \node (U) {\LARGE \faIcon{user}};

    \node[
        draw=gray,
        thick,
        text width=10em,
        align=center,
        cloud callout,
        cloud puffs=15,
        cloud puff arc=120,
        inner sep=-0.25em,
        callout pointer segments=2,
        callout relative pointer={(-9:0.8)},
        callout pointer start size=0.1 of callout,
        callout pointer end size=0.04 of callout,
        aspect=3,
        above left=-0.3 and 1.6 of U,
    ] (I) {I need some recipes for my new slow cooker...};

    \draw[
        decorate,
        decoration={
                text={|\footnotesize\itshape|User intent},
                text along path,
                text align=center,
            },
    ] ($(I.west) - (0, 0.15)$) to[out=-90,in=180] ($(I.south) - (0, 0.45)$);

    \node[
        draw=gray,
        rounded corners,
        inner sep=0.75em,
        below=0.5 of U,
    ] (Q) {slow cooking food \qquad \faIcon{search}};

    \draw[-latex] (U) -- (Q);

    \node[
        above=0 of Q.north west,
        anchor=south west,
    ] {\footnotesize \textit{Query}};

    \node[
        docheader,
        below left=1.25 and -1.5 of Q,
    ] (D-h-2) {Ingredients unsuitable for slow cooking};

    \node[
        docheader,
        left=2.5 of D-h-2.north,
        anchor=north,
    ] (D-h-1) {Ten easy slow cooker recipes};
    \node[
        docheader,
        right=2.5 of D-h-2.north,
        anchor=north,
    ] (D-h-3) {How long it takes to slow-cook beef};

    \foreach \D/\N in {1/3, 2/2, 3/2}
        {
            \foreach \i in {1, ..., \N}
                {
                    \coordinate (Ls) at ($(D-h-\D.south west) - (0, 0.1+\i*0.4)$);
                    \coordinate (Le) at (D-h-\D.south east |- Ls);
                    \draw[
                        faketext,
                    ] (Ls) -- (Le);
                }
            \begin{scope}[on background layer]
                \node[
                    fit=(D-h-\D)(Le),
                    draw,
                    fill=gray!15,
                    inner sep=0.75em,
                    outer sep=0.5em,
                    rounded corners,
                ] (D-\D) {};
            \end{scope}
        }

    \path (D-1.south west) -- (D-1.north west) node[
        midway,
        sloped,
        above=-0.1,
    ] {\footnotesize\textit{Documents}};

    \draw[
        -latex,
        qdarrow,
    ] (Q) to[
        out=-145,
        in=50,
    ] node[relevant] {} (D-1);

    \draw[
        -latex,
        qdarrow,
    ] (Q) to[
        out=-110,
        in=70,
    ] node[relevant] {} (D-2);

    \draw[
        -latex,
        qdarrow,
    ] (Q) to[
        out=-80,
        in=120,
    ] node[relevant] {} (D-3);

    \draw[
        -latex,
        thick,
    ] (I) to[
        out=-85,
        in=85,
    ] node[relevant] {} (D-1);
\end{tikzpicture}

%% file: rel_work.tex
\section{Related Work}
\label{sec:rel_work}
Several ranking datasets have been published that consider the concept of what we refer to as \emph{user intents}. Most notably, the data provided for the \trecweb{} track~\cite{collins2014trec} customarily includes topics (queries) along with \emph{topic descriptions} as well as, in many cases, \emph{subtopics}. These subtopics represent various distinct aspects that each topic may have. The data further includes relevance judgments for documents from the \clueweb{} collections w.r.t.\ the topics and subtopics. However, the \trecweb{} track has been discontinued after 2014, and \clueweb{} corpora are not freely available. Our dataset is similar, as the subtopics are essentially user intents.

The \ms{} ranking dataset~\cite{nguyen2016ms}, which has emerged as one of the most widely used collections for IR-related tasks in recent years, contains a large number of training and evaluation queries. Furthermore, the \trecdl{} track~\cite{craswell2020overview,craswell2021overview} provides annotated test sets of queries and corresponding relevance annotations. More recently, the second version of the \ms{} corpus, which is significantly larger than the first version, was released to be used in the \trecdl{} 2021 track and onward~\cite{craswell2022overview,craswell2023overview,craswell2024overview}.

\citet{mackie2021deep} showed that queries (topics) within \trecdl{} vary with respect to their complexity (and, hence, difficulty) and released the \dlhard{} dataset. Along with relevance annotations, this dataset assigns \emph{intent categories} to each query. Similarly, \emph{intent taxonomies} have been proposed for web search in general~\cite{cambazoglu2021intent} as well as legal case retrieval~\cite{shao2023intent}. The difference compared to our work is that we annotate specific user intents rather than categories.

Another related line of work deals with the \emph{reformulation} of complex queries. \citet{mackie2022codec} recently released the \codec{} collection for document and entity ranking, which also contains query reformulations. \citet{salamat2023neural} showed that the way queries are worded has an impact on their corresponding ranking performance. Our proposed user intents can be seen as reformulations that focus on specific aspects of the original query.

%% file: dataset_creation.tex
\section{The \name{} Dataset}
In this section, we introduce the \name{} dataset by outlining the creation and annotation process and presenting some statistics.

\subsection{Dataset Creation}
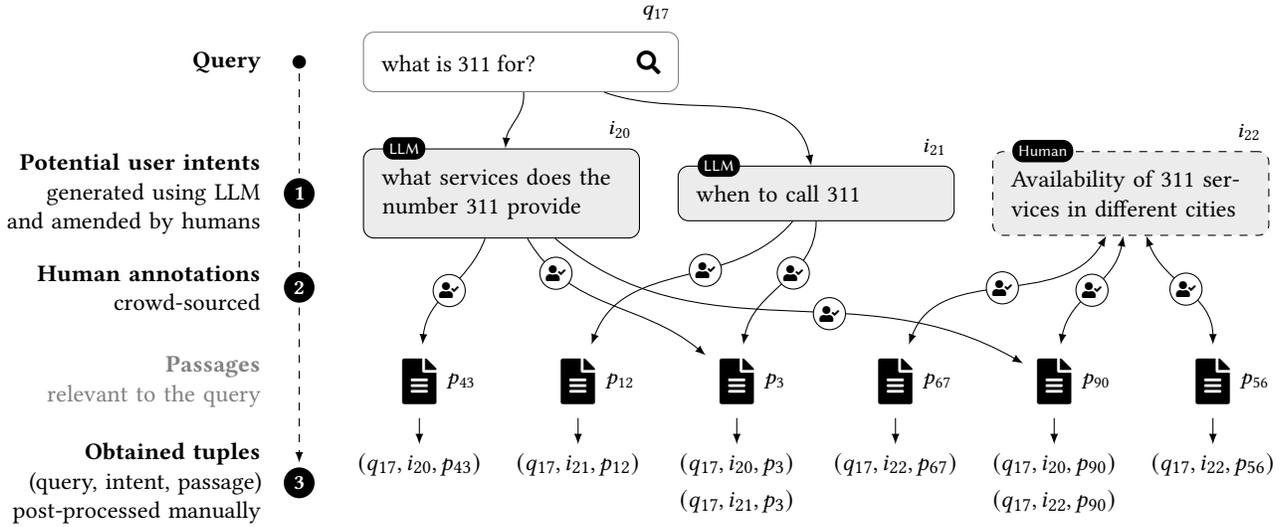
\begin{figure*}
    \centering
    \input{fig/approach}
    \caption{A high-level overview of how \name{} is created: Given a query, an LLM is used to generate candidate user intents. The query and its relevant passages (according to the original QRels), along with the candidate intents, are presented to human annotators, who can add, modify, or remove candidate intents and assign passages to them.}
    \label{fig:dataset.approach}
\end{figure*}
The process of creating the dataset comprises several key stages: generating candidate intents using an LLM (Section~\ref{subsec:generate_intent}), clustering and manual refinement of intents (Section~\ref{subsec:cluster_intent_selection}), crowd-sourcing annotations (Section~\ref{subsec:crowdsourcing}), merging similar intents (Section~\ref{subsec:intent_merging}) and QRel creation (Section~\ref{subsec:qrel_generation}). This process is illustrated in Fig.~\ref{fig:dataset.approach}.

\subsubsection{Generating Candidate User Intents}
\label{subsec:generate_intent}
For all queries in the \trecdl{}-21 and '22 test sets, we retrieve all relevant passages using their respective QRel files. We then cluster similar passages per query. To achieve this, we first obtain passage embeddings using Sentence-BERT~\cite{reimers2019sentence} and then group passages into the same cluster if their pairwise cosine similarity exceeds a threshold of $0.8$. In the next step, we select the query and passages from the clusters to give to the LLM to generate five distinct intents relevant to the query-passage pairs. We employ the GPT-4 model with the prompt given below. We use a temperature value of 0.6 to control randomness which helps in getting diverse intents.
\input{prompt-intents}


\subsubsection{Clustering and Intent Selection}
\label{subsec:cluster_intent_selection}
After generating intents, we cluster similar intents using the SBERT embedding and cosine similarity approach as described above. We group intents that are similar in meaning if their pairwise cosine similarity exceeds a threshold of $0.9$.
This clustering process helps in reducing redundancy and coming up with distinct intents.
After clustering, we do manual selection, where we examine the clustered intents and choose the most relevant ones for each cluster. 
We do this to remove irrelevant intents or hallucinated text by the LLM. 
If any intents are found to be incomplete or poorly written, they are manually rewritten to improve their clarity and comprehensiveness. 
This ensures that the intents are well-defined and useful for the next stages of the dataset creation process. After this process, only queries with 2 or more intents were selected which resulted in 26 queries. 

\subsubsection{Crowdsourcing Annotation}
\label{subsec:crowdsourcing}
The next step involves crowdsourcing to annotate the intents with the relevant passages. Our pool of annotators comprises volunteers who are computer scientists and graduate school students familiar with ranking tasks for search. Annotators are presented with a query and a passage and are asked to determine which of the provided intents the passage satisfies. Additionally, annotators are given the option to add or modify intents if they find that the existing ones do not capture the passage's intent. To manage queries with a large number of relevant passages (more than 30), the passages are divided into smaller chunks of 30. This division creates subqueries, making the annotation process more manageable for the annotators. Each subquery is annotated separately, ensuring that the workload is distributed and the annotators can focus on a smaller set of passages. In total, 22 sets of annotations are done by 16 distinct annotators and each set consist of 5 rounds (queries or subqueries), such that each query is annotated at least twice.

\subsubsection{Manual Review and Merging of Intents}
\label{subsec:intent_merging}
In order to improve data quality and avoid redundancy, we conduct a manual review and merge intents. We evaluate the intents suggested by the annotators and integrate them into the existing set of intents where appropriate. E.g., in Fig.~\ref{fig:dataset.approach} we merge "when to call 311" and "when to call 311 rather than 911" into a single intent. Any passage-intent pair which does not have at least two annotators is dropped to ensure that the final set of intents reflects a consensus among multiple annotators. The merging process also helps in consolidating similar intents and removing any redundant or less relevant ones. After this process, we end up with 24 queries. We further elaborate on different scenarios we encountered during this phase in Appendix \ref{appendix:merging_intents}

\subsubsection{Scoring and Creating QRel File}
\label{subsec:qrel_generation}
Finally, we score the intent-passage pairs and create a QRel file for ranking. The scoring is based on the annotations provided by the participants. Each intent-passage pair is scored as follows: a score of 0 is assigned if no annotator marked the intent, a score of 1 is assigned if at least one annotator marked the intent, and a score of 2 is assigned if all annotators marked the intent. These scores reflect the level of agreement among the annotators and the relevance of the intent to the passage. The final query-intent-passage-score mappings are compiled into a QRel file, which is used for ranking. This QRel file serves as ground-truth for evaluating information retrieval systems, ensuring that the dataset can be effectively used for further research and application.


\subsection{Statistics}
\label{subsec:stats}
\begin{figure}
    \begin{subfigure}{.49\linewidth}
        \centering
        \input{plot/stat_rel_1.tex}
        \caption{$\text{rel} \geq 1$}
    \end{subfigure}
    \begin{subfigure}{.49\linewidth}
        \centering
        \input{plot/stat_rel_2.tex}
        \caption{$\text{rel} = 2$}
    \end{subfigure}
    \caption{Histograms illustrating the number of relevant passages per intent for (a) all relevant passages and (b) only passages with relevance label $2$.}
    \label{fig:intent_pass_dist}
\end{figure}
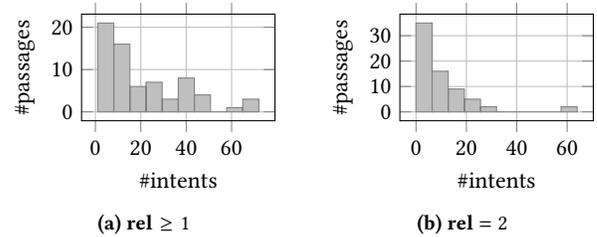

Initially, the dataset included 118 queries from \trecdl{}-21 and '22. Through a process of clustering and intent selection, 26 queries were identified as suitable for annotations, as these queries had two or more distinct intents (69 in total). 
After annotation (Sec.~\ref{subsec:crowdsourcing}), a manual review and merging of intents were performed (Sec.~\ref{subsec:intent_merging}). This process was necessary because the number of intents increased from 69 to 171 due to annotators adding custom intents. Hence, this review process was crucial in refining the dataset and ensuring the accuracy and clarity of the intents. After this rigorous review, 24 queries and 69 intents were finalized for inclusion in the dataset with \textbf{2655 relevance annotations} present in the final QRel file. The distribution of relevant passages per intent is shown in Fig.~\ref{fig:intent_pass_dist}.

Because annotators were able to add custom intents, computing established agreement measures is difficult as the intents annotated by humans may have different granularities; however, the relevance scores we obtained in Sec.~\ref{subsec:qrel_generation} are determined by the overlap of judgments and can therefore be seen as an indication of agreement among annotators.

\subsection{Tasks and Evaluation}
\label{subsec:tasks}
The \name{} dataset can be used for several tasks, such as:

\textbf{Intent-based ranking} aims at improving the document ranking by understanding different user intents and ensuring that the returned documents are relevant to the intent. This can be evaluated using metrics like nDCG@10.
    
\textbf{Diversity of search results} aims at ensuring that document rankings provide diverse sets of responses that cover various aspects of the query to satisfy users information needs, evaluated using metrics like $\alpha$-nDCG@10.

\textbf{Intent-based summarization} aims at generating a summary that covers multiple intents of a query, evaluated using metrics such as ROUGE or BLEU.

\textbf{User and machine intent alignment} aims at bridging the gap between user and machine intent through query rewriting to fully specify the intent~\cite{anand2023query}. \name{} aids in training generative models that can generate intents more aligned with real-world user intents.

%% file: fig/approach.tex
\tikzset{
    intent/.style={
            align=left,
            text width=10em,
            inner sep=0.8em,
            fill=gray!15,
            rounded corners,
            draw=black,
            label={
                    [fill=black, rounded corners, font=\sffamily\scriptsize, text=white, anchor=west, xshift=-5em, inner sep=0.25em]above:#1
                },
        },
    annotation/.style={
            circle,
            draw=black,
            fill=white,
            inner sep=0.1em,
            font=\scriptsize,
            execute at begin node=\faIcon{user-check},
        },
    desc/.style={
            align=right,
            text width=11em,
        },
    identifier/.style={
            font=\small,
        },
    stepnum/.style={
            circle,
            fill=black,
            text=white,
            font=\small\bfseries,
            inner sep=0.2em,
            outer sep=0.2em,
        }
}

\begin{tikzpicture}
    \node[
        draw=gray,
        rounded corners,
        inner sep=0.75em,
    ] (Q) {what is 311 for? \qquad\qquad \faIcon{search}};

    \node[
        desc,
        left=1.25 of Q,
    ] (Q-d) {\textbf{Query}};
    \node[
        stepnum,
        right=0.25 of Q-d,
    ] (Q-d-s) {};

    \node[
        identifier,
        above=0.25 of Q.north east,
        anchor=east,
    ] {$q_{17}$};

    \node[
        intent={LLM},
        below=1.75 of Q.west,
        anchor=west,
    ] (I-1) {what services does the number 311 provide};
    \node[
        identifier,
        above=0.25 of I-1.north east,
        anchor=east,
    ] {$i_{20}$};
    \draw[-latex] (Q) to[
        out=275,
        in=85,
    ] (I-1);

    \node[
        intent={LLM},
        right=0.5 of I-1,
    ] (I-2) {when to call 311};
    \node[
        identifier,
        above=0.25 of I-2.north east,
        anchor=east,
    ] {$i_{21}$};
    \draw[-latex] (Q) to[
        out=340,
        in=100,
    ] (I-2);

    \node[
        intent={Human},
        right=0.5 of I-2,
        dashed,
    ] (I-3) {Availability of 311 services in different cities};
    \node[
        identifier,
        above=0.25 of I-3.north east,
        anchor=east,
    ] {$i_{22}$};

    \node[
        desc,
    ] (I-d) at (Q-d |- I-1) {\textbf{Potential user intents}\\generated using LLM\\and amended by humans};
    \node[
        stepnum,
    ] (I-d-s) at (Q-d-s |- I-d) {1};

    \foreach \p [count=\idx] in {43, 12, 3, 67, 90, 56}
        {
            \coordinate (L) at ($(I-1.west) + (0.75, 0)$);
            \coordinate (R) at ($(I-3.east) + (-0.75, 0)$);
            \coordinate (C) at ($(L)!1/5*(\idx-1)!(R) + (0, -2.5)$);
            \node[
                outer sep=0.25em,
            ] (P-\idx) at (C) {\Huge \faIcon{file-alt}};
            \node[
                identifier,
                right=-0.15 of P-\idx,
            ] (P-i-\idx) {$p_{\p}$};
        }

    \coordinate (Temp) at ($(P-1)!0.5!(P-i-1)$);
    \node[
        desc,
        text=gray,
    ] (P-d) at (Temp -| I-d) {\textbf{Passages}\\relevant to the query};

    \node[
        desc,
    ] (C-d) at ($(I-d)!0.5!(P-d)$) {\textbf{Human annotations}\\crowd-sourced};
    \node[
        stepnum,
    ] (C-d-s) at (Q-d-s |- C-d) {2};

    \draw[-latex] (I-1) to[
        out=250,
        in=85,
    ] node[annotation] {} (P-1);
    \draw[-latex] (I-1) to[
        out=300,
        in=140,
    ] node[annotation, pos=0.2] {} (P-3);
    \draw[-latex] (I-1) to[
        out=320,
        in=150,
    ] node[annotation, pos=0.6] {} (P-5);

    \draw[-latex] (I-2) to[
        out=230,
        in=70,
    ] node[annotation, pos=0.4] {} (P-2);
    \draw[-latex] (I-2) to[
        out=270,
        in=80,
    ] node[annotation, pos=0.4] {} (P-3);

    \draw[latex-latex] (I-3) to[
        out=240,
        in=70,
    ] node[annotation] {} (P-4);
    \draw[latex-latex] (I-3) to[
        out=260,
        in=80,
    ] node[annotation] {} (P-5);
    \draw[latex-latex] (I-3) to[
        out=290,
        in=90,
    ] node[annotation] {} (P-6);

    \foreach \I/\p [count=\idx] in {{20}/43, {21}/12, {20,21}/3, {22}/67, {20,22}/90, {22}/56}
        {
            \foreach \i [count=\n] in \I
            {
                \coordinate (T) at ($(P-\idx) + (0, -0.5*\n-0.6)$);
                \node (T-\idx-\n) at (T) {$(q_{17}, i_{\i}, p_{\p})$};
            }
            \draw[-latex] (P-\idx) -- (T-\idx-1);
        }

    \coordinate (Temp) at ($(T-3-1)!0.5!(T-3-2)$);
    \node[
        desc,
    ] (T-d) at (P-d |- Temp) {\textbf{Obtained tuples}\\(query, intent, passage)\\post-processed manually};
    \node[
        stepnum,
    ] (T-d-s) at (Q-d-s |- T-d) {3};

    \draw[
        -latex,
        dashed,
    ] (Q-d-s) -- (I-d-s) -- (C-d-s) -- (T-d-s);
\end{tikzpicture}

%% file: prompt-intents.tex
\begin{tcolorbox}[title=\textbf{LLM Prompt}: Intent candidate generation]
\small
A person wants to find out distinct intention behind the question \textbf{\{query\}}. Give five descriptive (max. 15 words) distinct intentions which are easy to understand. Consider all documents in your response. Response should be in this format:

\textbf{Intention}:: <intention> , \textbf{Doc\_list}::<list of documents with the intention> \\ \textbf{Documents}: \textbf{\{list of input documents\}}
\end{tcolorbox}

%% file: plot/stat_rel_1.tex
\begin{tikzpicture}
    \begin{axis}[
            ybar,
            height=3cm,
            width=\linewidth,
            xlabel={\#intents},
            ylabel={\#passages},
            grid=major,
        ]
        \addplot+[
            hist={
                    bins=10,
                },
            fill=lightgray,
            draw=gray,
        ] table[
                col sep=comma,
                y=rel_1,
            ] {plot/psg_per_intent.csv};
    \end{axis}
\end{tikzpicture}

%% file: plot/stat_rel_2.tex
\begin{tikzpicture}
    \begin{axis}[
            ybar,
            height=3cm,
            width=\linewidth,
            xlabel={\#intents},
            ylabel={\#passages},
            grid=major,
        ]
        \addplot+[
            hist={
                    bins=10,
                },
            fill=lightgray,
            draw=gray,
        ] table[
                col sep=comma,
                y=rel_2,
            ] {plot/psg_per_intent.csv};
    \end{axis}
\end{tikzpicture}

%% file: experimental_setup.tex
\section{Experiments}
\label{sec:experiment}
In order to demonstrate the utility of \name{}, we conduct experiments using a number of simple baselines:
\textbf{\bm{}}~\cite{bm25} is a lexical model which is also used as a first-stage retriever for re-rankers. 
\textbf{\bert{}}~\cite{devlin_bert_2018} is a cross-attention re-ranker (\bert{}-base, 12 layers). The input length is restricted to a maximum of 512 tokens. The model is trained on \ms{} passage data using a pointwise ranking loss objective with a learning rate of 1e-5.
\textbf{\col{}} \cite{santhanam-etal-2022-colbertv2} is a multi-vector late-interaction re-ranking model that computes token-wise representations for the query and document and estimates relevance using the MaxSim operation.



\subsection{Results}
\begin{table}
        \centering
        \begin{tabular}{lcc}
                \toprule
                        & \textbf{Intent Ranking} & \textbf{Diversity}                              \\
                \cmidrule(lr){2-2} \cmidrule(lr){3-3}
                        & nDCG@10        & $\alpha$-nDCG@10                                  \\
                \midrule
                \multicolumn{3}{l}{\small\textbf{\textsf{Original queries, user intent QRels}}}   \\
                \obm{}    & 0.073          & 0.144                                             \\
                \bert{} & 0.060          & 0.114                                             \\
                \midrule
                \multicolumn{3}{l}{\small\textbf{\textsf{User intents as queries, user intent QRels}}} \\
                \obm{}    & 0.116          & 0.250                                             \\
                \bert{} & 0.169          & 0.375                                            \\
                \col{}  & \textbf{0.261} & \textbf{0.532}                                    \\
                \bottomrule
        \end{tabular}
        \caption{\name{} ranking performance. Best performing models are in bold. Re-rankers use the corresponding \obm{} runs. Diversity is calculated at query level in both cases.}
        \label{tab:results}
\end{table}
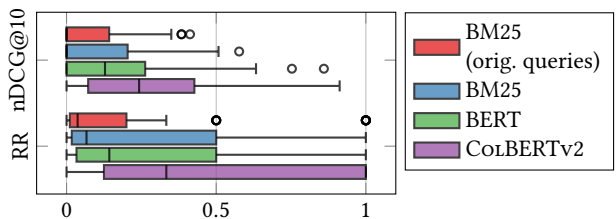
\begin{figure}
    \centering
    \input{plot/perf_cmp.tex}
    \caption{Performance comparison on a per-intent level. The boxplots show the distribution of the ranking performance of individual intents.}
    \label{fig:ranking_perf_cmp}
\end{figure}




We report results on two of the tasks outlined in Section~\ref{subsec:tasks}, namely \textit{intent-based ranking} and \textit{diversity of search results}. We present these results in Table~\ref{tab:results}. Note that we evaluate two settings: First, we use the original queries, but evaluate using the user intent-based QRels (i.e., assuming that the user had one specific intent in mind). Second, we treat the user intents as queries directly. The results show that, unsurprisingly, specifying the actual user intent as the query results in better performance than using the (more general) original queries. We additionally demonstrate the diversity ranking performance of various models using the $\alpha$-nDCG@10 metric. To achieve this in the second setting (where user intents are treated as queries), we employ reciprocal rank fusion~\cite{cormack2009reciprocal} with $k=60$. This technique is applied to the intent-based rankings to generate a unified ranking for the original query. Overall, \col{} shows the best performance. 
Finally, we closely examine the ranking performance corresponding to each user intent in Fig.~\ref{fig:ranking_perf_cmp}. The results are in line with Table~\ref{tab:results}.

The key takeaway from these results is the necessity of specifying concrete user intents; in other words: if a user has a specific information need, it is necessary to provide that intent as a clear, unambiguous query to a search engine.

%% file: plot/perf_cmp.tex
\begin{tikzpicture}[every mark/.append style={mark size=1.5pt}]
    \begin{axis}[
            width=0.75\linewidth,
            height=4cm,
            xmajorgrids,
            ytick={1.5, 6.5},
            yticklabels={RR, nDCG@10},
            yticklabel style={rotate=90},
            cycle list={
                    {draw=black, fill=c1, solid, thick, opacity=0.75, mark=o},
                    {draw=black, fill=c2, solid, thick, opacity=0.75, mark=o},
                    {draw=black, fill=c3, solid, thick, opacity=0.75, mark=o},
                    {draw=black, fill=c4, solid, thick, opacity=0.75, mark=o}
                },
            area legend,
            legend entries={\obm{}\\(orig.\ queries), \obm{}, \bert{}, \col{}},
            legend cell align={left},
            legend style={
                    at={(1.025, 1)},
                    anchor=north west,
                    cells={align=left},
                },
        ]

        \addplot+[
            boxplot={draw position=3},
        ] table[
                y=rr,
                col sep=comma,
            ] {plot/perf_bm25_queries.csv};
        \addplot+[
            boxplot={draw position=2},
        ] table[
                y=rr,
                col sep=comma,
            ] {plot/perf_bm25_intents.csv};
        \addplot+[
            boxplot={draw position=1},
        ] table[
                y=rr,
                col sep=comma,
            ] {plot/perf_bert.csv};
        \addplot+[
            boxplot={draw position=0},
        ] table[
                y=rr,
                col sep=comma,
            ] {plot/perf_colbert.csv};

        \addplot+[
            boxplot={draw position=8},
        ] table[
                y=ndcg_10,
                col sep=comma,
            ] {plot/perf_bm25_queries.csv};
        \addplot+[
            boxplot={draw position=7},
        ] table[
                y=ndcg_10,
                col sep=comma,
            ] {plot/perf_bm25_intents.csv};
        \addplot+[
            boxplot={draw position=6},
        ] table[
                y=ndcg_10,
                col sep=comma,
            ] {plot/perf_bert.csv};
        \addplot+[
            boxplot={draw position=5},
        ] table[
                y=ndcg_10,
                col sep=comma,
            ]{plot/perf_colbert.csv};
    \end{axis}
\end{tikzpicture}

%% file: conclusion.tex
\section{Conclusion}
In this paper, we have created the \name{} dataset to understand user intents, thereby satisfying information needs more effectively. We have used queries from \trecdl{}-21 and \trecdl{}-22, generated intents using an LLM, and crowd-sourced relevance annotations. \name{} can be used for a variety of tasks; we present performance of different models on ranking and diversity tasks, showing the importance of this dataset for fulfilling user information needs.
For future work, we plan to extend \name{} to include queries from \trecdl{}-19, \trecdl{}-20, and \dlhard{}.

%% file: appendix.tex
\appendix
\section{Generating Intents using LLM}
\label{apndx:llm_gen}
\input{table/appendix_intent}
The objective of this step is to generate a diverse set of user intents that accurately reflect the informational needs expressed by the user query. To achieve this, we first retrieve all relevant passages for each query from the QRel file. We observed a significant amount of redundancy among these passages, which could lead to duplication in intent generation when using a large language model (LLM). To mitigate this, we cluster similar relevant passages for each query before proceeding with intent generation. Several methods were explored, including entailment-based approaches , but we found that clustering using cosine similarity with Sentence-BERT~\cite{reimers2019sentence} (detailed in Appendix~\ref{apndx:intent_clust}) yielded the best results. For the query "What is 311 for" in Table~\ref{tab:appen_intent}, there are 53 relevant passages. After applying passage clustering, these were reduced to 18 distinct clusters. Next we give the query and 18 passages too LLM for intent generation. Subsequently, we generated intents using an LLM, resulting in 10 intents, as shown in Table~\ref{tab:appen_intent}.

We experimented with various prompts (some examples are shown in Fig below) following the Context-Aware Query Expansion (CAQE) method~\cite{anand2023context}, which generates intents based on a query and a relevant document/passage pair. However, this approach resulted in a large number of intents, many of which were duplicates, as they were derived from individual \textit{<query, passage>} pairs. To address this issue, we expanded the context to include a list of passages, which allowed us to generate a smaller number of high-quality intents by considering the collective context of all relevant passages (prompt in Section~\ref{subsec:generate_intent}).

\begin{tcolorbox}[title=\textbf{LLM Prompt using CAQE}: Intent generation]
\small
\textbf{Prompt 1:} Given a query and document generate the intention of the query given the document.  The generation should be like a human would write an intention in more than 3 words and less than 10 words. Use 'UNKNOWN' if there is no intent found. Response should as human like with minimum 3 words and maximum 10 words and not the answer to the query, only intention. 
\textbf{query}:{\{query\}} 
\textbf{document}:{\{doc\}} \\

\textbf{Prompt 2:} Being a intent generator, the task is to generate multiple intents of min 3 words. Given a query and a document, generate multiple distinct intents from the document that answers the query. Below are the rules to be followed. Answer in one short sentence per intent of minimum 3 words. Generate only distinct intents and not answers.  Make sure intent is not the Query. Generate multiple intents but limit to maximum 3. Use 'UNKNOWN' if there is no intent found. Response should be in this specific format Query:: <query>
Query\_Intent:: <intent>.
\textbf{query}:{\{query\}} 
\textbf{document}:{\{doc\}} \\

\end{tcolorbox}

\section{Intent Clustering}
\label{apndx:intent_clust}
As outlined in Section~\ref{apndx:llm_gen}, we perform clustering both before (passage clustering) and after (intent clustering) intent generation using a large language model (LLM) to eliminate redundancy. The same clustering approach is applied in both stages. For a given query, we first obtain embeddings for all passages or intents using Sentence-BERT (SBERT). Next, we select a passage/intent $P_i$/$I_i$ and identify all other passages/intents that have a pairwise cosine similarity above a predefined threshold. Specifically, we use a threshold of $0.8$ for passages and $0.9$ for intents. All passages/intents that meet this similarity criterion are grouped into the same cluster, and then removed from the passage/intent list. This process is repeated iteratively until no passages/intents remain in the list. For the query "What is 311 for," we generated intents, which are listed in the "LLM generated intents" section of Table~\ref{tab:appen_intent}. Upon clustering these intents, we obtained two clusters: one consisting of 9 intents and another consisting of a single intent. Next we select intent representative of the cluster and reformulate it for the next step. So "differentiate between emergency and non-emergency numbers" is reformulated to "when to call 311" and representative reformulated intent from cluster 1 is "what services does the number 311 provide".

In addition to using SBERT with cosine similarity, we experimented with alternative methods for similarity scoring, including out-of-the-box and fine-tuned entailment models with both uni-directional and bi-directional entailment. To assess the quality of the clustering, we constructed an evaluation set on which all the clustering methods, for both passages and intents, were systematically evaluated. SBERT-based clustering approach performed better than the alternative methods. 

\section{Annotation via Crowdsourcing}
\label{apndx:annotation}
\begin{figure*}
    \centering
    \frame{\includegraphics[height=.9\textheight]{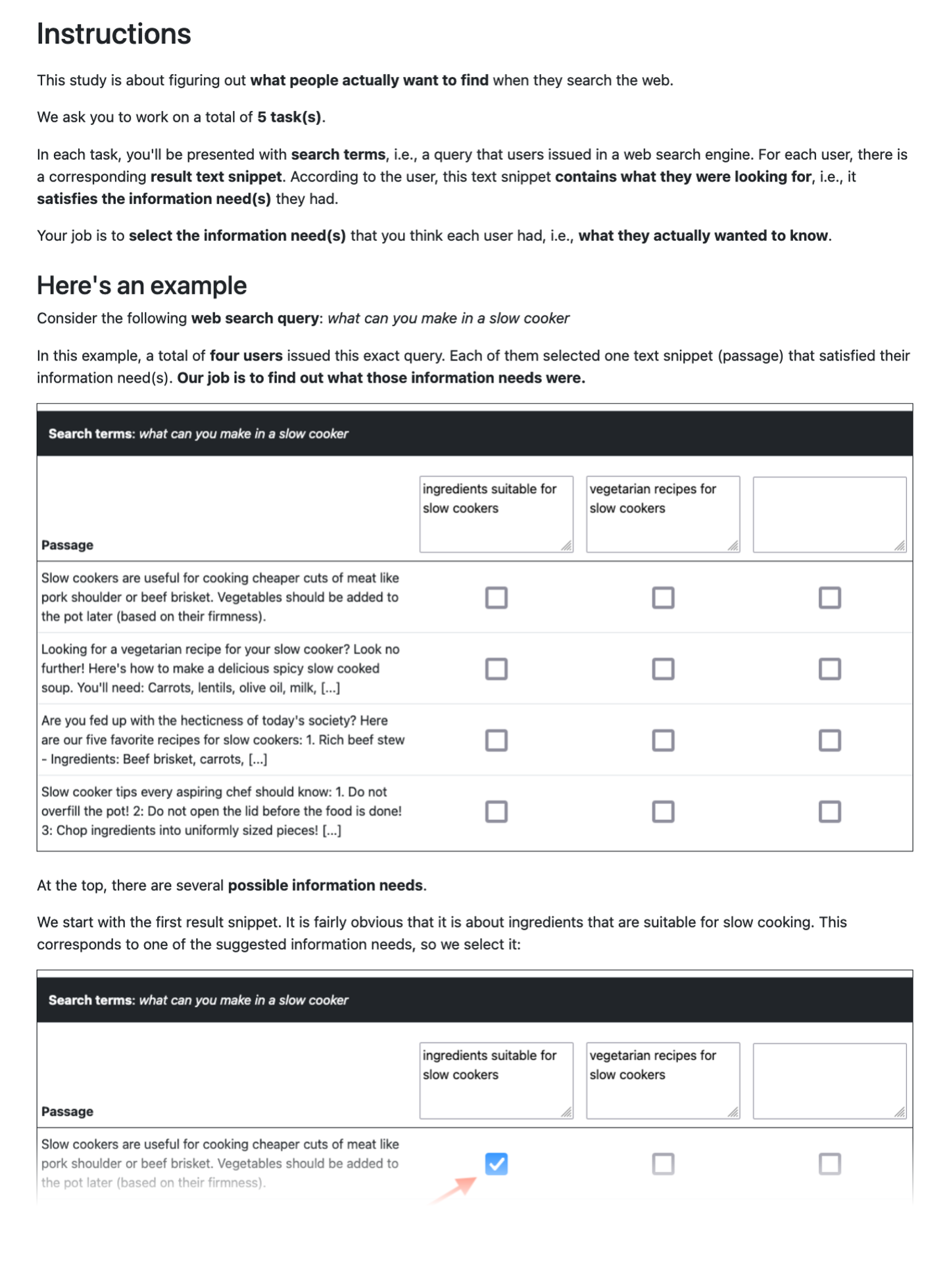}}
    \caption{The instructions displayed to each crowdsourcing worker prior to the annotation process. Note that this screenshot is cropped and does not include the entire instructions.}
    \label{fig:apndx:annotation.instructions}
\end{figure*}
\begin{figure*}
    \centering
    \frame{\includegraphics[width=\textwidth]{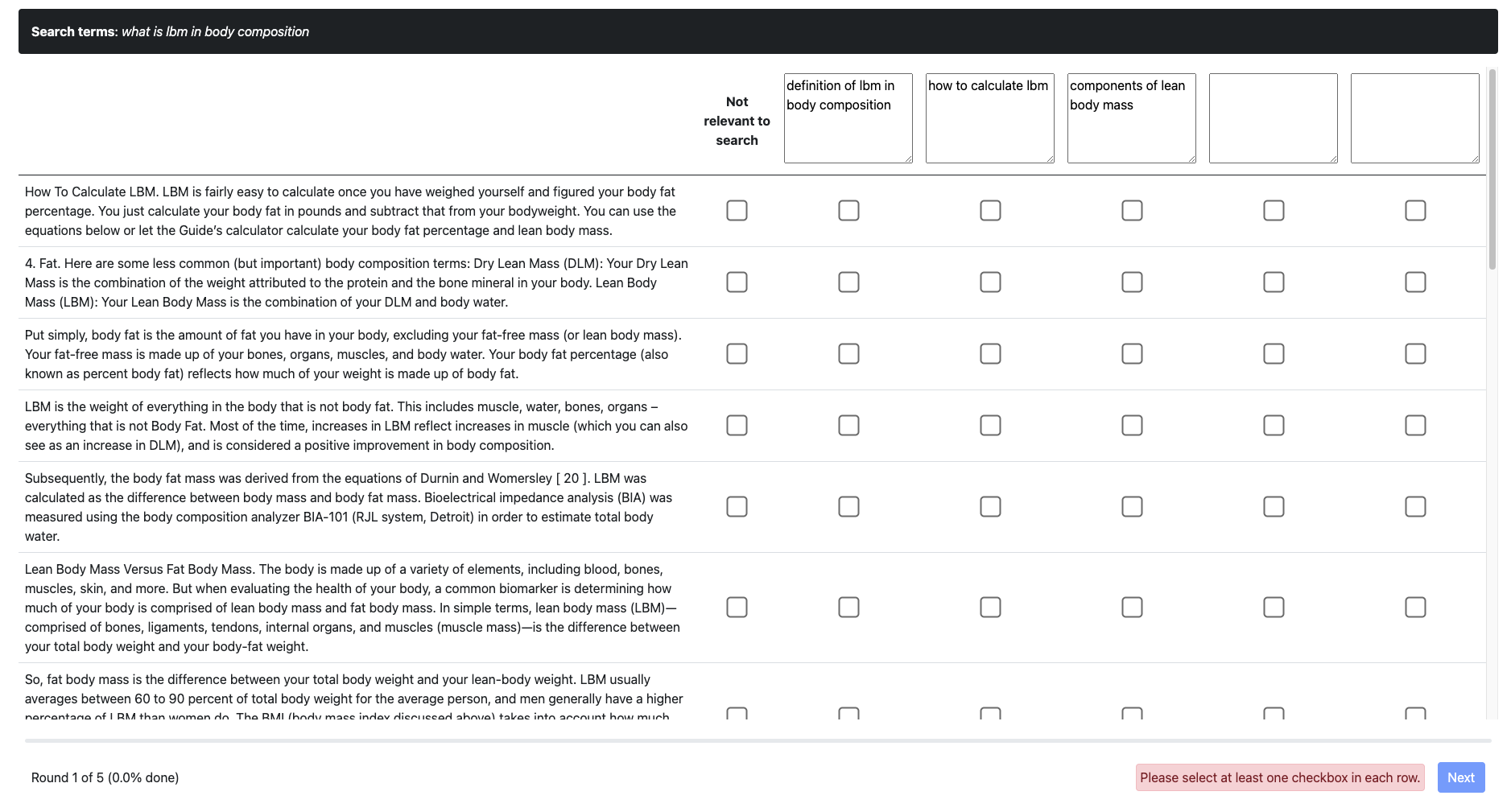}}
    \caption{The user intent annotation interface for crowdsourcing workers. Each participant is asked to complete several \emph{rounds}, where a round corresponds to one (sub)query and the corresponding list of passages to annotate. The page presents the original query (\emph{search terms}) and the intent candidates (generated by the LLM) to the participant. At the top, the suggested intents can be modified and new intents can be added. Alternatively, it is possible to indicate that a given passage is not relevant to the search query at all.}
    \label{fig:apndx:annotation.interface}
\end{figure*}
The collection of user intent annotations for \name{} is performed using a custom web application we implemented using the \otree{} framework~\cite{chen2016otree}. 

Each participant is presented with detailed instructions how to perform the task (cf.\ Fig.~\ref{fig:apndx:annotation.instructions}) in the beginning. The subsequent pages display \textbf{one (sub)query each} along with a list of the corresponding passages and intent candidates (cf.\ Fig.~\ref{fig:apndx:annotation.interface}). The interface ensures that each passage either has at least one annotated relevant intent or it is specifically indicated that the passage in question is not relevant to the query at all. By displaying all passages and intents within the same page we make sure that the participant always maintains a mental overview over the distinct intents (i.e., aspects) of the current query.

The collected data is stored in a PostgreSQL database. After the collection is complete, \otree{} provides functionality to export the relevant data in CSV format. Our web application for data collection is publicly available.

\section{Cleanup and Merging of Intents Post-Annotation}
\label{appendix:merging_intents}
After the annotation phase, we perform a manual analysis of intents and corresponding relevance annotations for the passages. We show some of the scenarios and related intents in Table \ref{tab:merging_examples}. We observe from example one that Intent 2 and Intent 3 are semantically similar and can be considered as redundant and hence are merged. When merging the intents, we also combine the ratings suing the following guidelines:
\begin{itemize}
\item We set the score based on number of annotations for the intents to be merged that indicate the level of agreement between annotators and relevance to the query
    \item For instance, if two annotators annotate 1 for both intents paired with a certain passage, we assign the relevance score as 2 for the merged intent.
    \item If only one annotator assigned a score of 1 to both intents for the same passage we assign the score as 1.
\end{itemize}

Apart from redundant intents, we also observed cases where the machine generated or the custom intents from the user were not relevant to the core aspects of the query. For instance, in example 2 in Table \ref{tab:merging_examples}, ``the cost of Tuk-tuks'' is irrelevant to the query which deals with aspects related to cost of living in Bangkok. Such intents are removed. 

We also observe cases where the intents are same as query as shown in the table and these intents are removed. Finally, we also observe a scenario where the generated intents answer the query instead of conveying the explicit or latent aspects of the query as shown in Example 4 in Table \ref{tab:merging_examples}. We remove such instances as they are actually not intents.
\input{table/merging_intents_samples}

%% file: table/appendix_intent.tex
\begin{table*}
\begin{tabular}{p{0.15\linewidth} | p{0.7\linewidth} | p{0.05\linewidth}} 
\hline
\multicolumn{3}{|c|}{\textbf{Query: What is 311 for}} \\
\hline
\multicolumn{3}{|c|}{\textbf{LLM generated intents (Section~\ref{subsec:generate_intent})}} \\
\hline
LLM generated intents & To identify 311 as a number for non-emergency law enforcement related complaints, 
\newline to explain 311 as access to various non-emergency municipal services, 
\newline to highlight 311 for city information and non-emergency service requests, 
\newline to outline the general usage of 311 for non-emergency information and services, 
\newline to showcase the origin and adoption of 311 in various cities, 
\newline understand the utility of 311 as a contact number, 
\newline learn about 311's role in specific cities or counties, 
\newline discover how to report specific issues with 311, 
\newline find out about the origin and development of the 311 system, 
\newline differentiate between emergency and non-emergency numbers. & 10\\
\hline
\multicolumn{3}{|c|}{\textbf{Clustering (Section~\ref{subsec:cluster_intent_selection})}} \\
\hline
Cluster 1               & to identify 311 as a number for non-emergency law enforcement related complaints.,
\newline to explain 311 as access to various non-emergency municipal services.,
\newline to highlight 311 for city information and non-emergency service requests.,
\newline to outline the general usage of 311 for non-emergency information and services.,
\newline to showcase the origin and adoption of 311 in various cities.,
\newline understand the utility of 311 as a contact number.,
\newline learn about 311's role in specific cities or counties.,
\newline discover how to report specific issues with 311.,
\newline find out about the origin and development of the 311 system. & 9 \\
\hline
Cluster 2                & differentiate between emergency and non-emergency numbers. & 1 \\

\hline
\multicolumn{3}{|c|}{\textbf{Crowdsourcing Intents (Section~\ref{subsec:crowdsourcing})}} \\
\hline
Intents shown to Annotators & what services does the number 311 provide, 
\newline when to call 311 & 2\\
\hline
Annotator generated intents & What services are provided by 311 in different cities, 
\newline Availability of 311 system in city, 
\newline Who answers the call when 311 is dialed, 
\newline Availability of 311 services in different cities, 
\newline when to call 311 rather than 911, 
\newline what happens when one dials 311 & 6\\
\hline
\multicolumn{3}{|c|}{\textbf{Final Intents (Section~\ref{subsec:intent_merging})}} \\
\hline
Final Intents & when to call 311, 
\newline Availability of 311 services in different cities & 2\\    
\hline

\end{tabular}
\caption{The table illustrates the various stages in refining LLM-generated intents to final intents for a given query. The right column displays the number of intents at each stage.}
\label{tab:appen_intent}
\end{table*}

%% file: table/merging_intents_samples.tex
\begin{table*}
\begin{tabular}{lp{.68\textwidth}p{2cm}}
\toprule
    \textbf{Type} & \textbf{Query with Intents} & Decision \\
\midrule
\small

Redundant Intents  & \textbf{Query}: what vaccination should u give show piglets?
\\ & \textbf{Intent 1:} available vaccinations for show piglets \\
 & \textbf{Intent 2:} \textcolor{red}{optional vaccinations for show piglets} & Merge intent 2 \\
  & \textbf{Intent 3:} \textcolor{red}{non-essential vaccination for show piglets} & and intent 3  \\

\midrule
 Intents  & \textbf{Query}: How much money do i need in bangkok?
\\not relevant to query & \textbf{Intent 1:} how expensive is daily life in bangkok \\
 & \textbf{Intent 2:} how expensive is tourism in bangkok & Remove Intent 4 \\ 
 & \dots \\
  & \textbf{Intent 4:} \textcolor{red}{cost of taxi/tuk-tuks}  \\ 
  \midrule   & \textbf{Query}: when a house goes into foreclosure what happens to items on the premises?
\\ Same as query & \textbf{Intent 1:} what happens to personal items when a house goes into foreclosure? \\
 & \textbf{Intent 2:} what happens to fixtures when a house goes into foreclosure & Remove Intent 8 \\ 
 & \dots \\
  & \textbf{Intent 8:} \textcolor{blue}{what physically happens to items after a house goes into foreclosure?}  \\
\midrule
  Answers the query & \textbf{Query:} what is the name of the triangular region at the base of the bladder? & \\
  & \textbf{Intent 1:} \textcolor{purple}{Description of the trigone region?} & Remove Intent 1 and intent 2\\
 & \textbf{Intent 2:} \textcolor{purple}{Synonyms of the term trigone in bladder} & \\
  
  \hline
\end{tabular}
\caption{Examples of different scenarios for post-cleanup or merging of intents.}
\label{tab:merging_examples}
\end{table*}